\documentclass{article}

\usepackage{arxiv}

\usepackage[utf8]{inputenc} 
\usepackage[T1]{fontenc}    
\usepackage{hyperref}       
\usepackage{url}            
\usepackage{booktabs}       
\usepackage{amsfonts}       
\usepackage{nicefrac}       
\usepackage{microtype}      
\usepackage{lipsum}		
\usepackage{graphicx}
\usepackage{gensymb}
\usepackage{doi}
\usepackage[numbers]{natbib}

\title{A SIMPLE and COMPACT PASSIVE RESONANT FIBER OPTIC GYROSCOPE by USING NON-RECIPROCAL POLARIZATION TECHNIQUES}

\author{
  Onder Akcaalan\\
  Independent Researcher \\
  Hamburg, Germany\\
  \texttt{\ onderakcaalan@gmail.com*} \\
     \And
 Melike Gumus Akcaalan\\
  Institute for Nanostructure and Solid-State Physics\\
  University of Hamburg \\
  Hamburg, Germany\\
  \texttt{\ melike.gumus.akcaalan@uni-hamburg.de} \\
}

\renewcommand{\shorttitle}{\textit{arXiv} Template}

\hypersetup{
pdftitle={A template for the arxiv style},
pdfsubject={q-bio.NC, q-bio.QM},
pdfauthor={David S.~Hippocampus, Elias D.~Striatum},
pdfkeywords={First keyword, Second keyword, More},
}

\begin{document}
\maketitle

\begin{abstract}
We present a novel passive resonant fiber-optic gyroscope (RFOG) design that achieves two true quadrature points at $\pi/2$ and $3\pi/2$, enabling angular rotation measurement with maximum sensitivity. The use of a broadband light source, as demonstrated in previous studies, eliminates the need for precise frequency locking, while the resonant enhancement allows high sensitivity with significantly shorter fiber lengths. Building on this approach, the present work integrates a Non-Reciprocal Polarization-Dependent Phase Shifter (NRPPS) into the broadband RFOG configuration, called NRPPS-RFOG, enabling truly passive operation without the need for active modulation–demodulation. Theoretical analysis demonstrates that the coupling ratio and NRPPS phase shift critically influence both sensitivity and linearity. An additional NRPPS section extends the measurable angular rotation range, addressing limitations arising from long fiber loops. The proposed design combines compactness, high performance, and low complexity, offering a practical path toward navigation-grade RFOGs. These results highlight the potential of passive NRPPS-RFOGs for cost-effective, high-precision rotational sensing in demanding applications.

\end{abstract}


\section{Introduction}

Fiber-optic gyroscopes (FOGs) represent one of the most successful implementations of optical gyroscope technology, offering high accuracy \cite{sanders1996fiber}, wide dynamic range \cite{ono1991phase}, and long-term reliability \cite{korkishko2012interferometric} for navigation and guidance applications across commercial and military domains, including aerospace, maritime, defense, and autonomous vehicles. Operating on the Sagnac effect \cite{sagnac1913ether}, FOGs measure rotation by detecting the phase difference accumulated between light waves propagating in opposite directions within a fiber coil. Two primary configurations are commonly employed: interferometric fiber-optic gyroscopes (IFOGs) and resonant fiber-optic gyroscopes (RFOGs).

IFOGs have been widely adopted due to their tactical, navigational, and strategic capabilities. Achieving high sensitivity in IFOGs typically requires long fiber lengths or large coil diameters. However, extended fiber coils introduce several challenges, including thermal-induced phase noise (TPN) \cite{takei2023simultaneous,song2017modeling}, intensity-dependent nonlinearities (Kerr effect) \cite{bergh1982source}, polarization non-reciprocity \cite{chamoun2015noise}, optical losses \cite{lefevre2013fiber}, back-scattering, Rayleigh scattering \cite{lloyd2013modeling}, and limitations in coil size and packaging. These factors not only degrade performance but also long fiber and/or larger diameter spool cause complication of system integration.

To address these limitations, RFOGs were proposed as a compact alternative. Unlike IFOGs, RFOGs exploit the resonance of the fiber cavity, allowing light to circulate multiple times and effectively increasing the optical path length \cite{carroll1987passive,iwatsuki1984effect,watsuki1986eigenstate}. This resonant enhancement enables higher sensitivity with significantly shorter fiber lengths. Despite their theoretical advantages, conventional RFOGs rely on narrow-linewidth lasers, precise frequency locking \cite{sanders2018improvements}, and active modulation-demodulation schemes \cite{li2020signal} to determine rotation direction, which increases system complexity and cost.

A significant achievement was recently demonstrated by Zhao et al. \cite{zhao2022navigation}, who developed a navigation-grade RFOG using an ultra-simple white-light multi-beam interferometry approach. By employing a broadband light source instead of a narrow-linewidth laser, the need for complex frequency locking was eliminated, achieving bias instability and angular random walk at navigation-grade levels. While this approach substantially simplifies the system, active modulation-demodulation remains necessary. To overcome this limitation, Ovchinnikov and colleagues proposed a passive RFOG design based on a 3×3 fiber coupler \cite{ovchinnikov2023prototype}. However, due to the inherent design of the 3×3 coupler, the quadrature point deviates from $\pi/2$ and  preventing the reaching of maximum sensitivity.

In this work, we propose— to the best of our knowledge— the first passive RFOG architecture that achieves two true quadrature operating points at $\pi/2$ and $3\pi/2$. By exploiting these quadrature points, the angular rotation can be determined with maximum sensitivity. This approach, initially developed for IFOGs, has been shown to theoretically increase sensitivity by up to 40 times depending on fiber length \cite{akcaalan2025phase,akcaalan2025simple}, and it is thought that this design can also be used for RFOGs, allowing the system to be operated passively without the need for active modulation/demodulation. In this light of thought, the paper, first, will introduce the simplest broadband RFOG configuration. Next, a passive Non-Reciprocal Polarization-Dependent Phase Shifter (NRPPS), previously applied in IFOG designs to eliminate modulation/demodulation provided by the MIOC, is incorporated into the RFOG system with a detailed theoretical analysis. Finally, we investigate how key parameters including the coupling ratio and the NRPPS phase shift values, influence the sensitivity of the proposed design.

\begin{figure}[!t]
\centering
\includegraphics[width=0.95\textwidth]{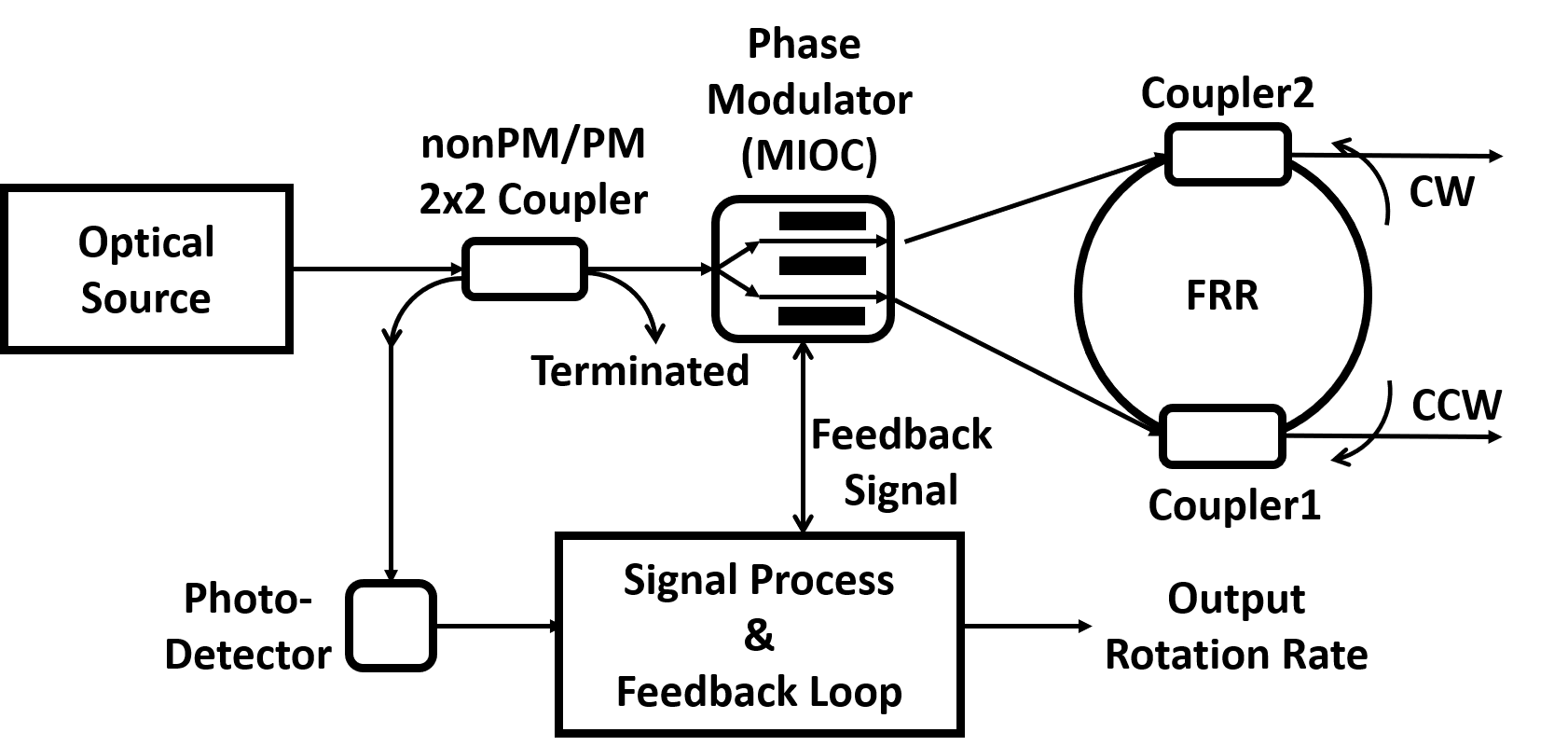}
\caption{\label{fig:RFOG_Setup} The design of the RFOG using ultra-simple white-light multibeam interferometry \cite{zhao2022navigation}}
\end{figure}

\begin{figure}[!b]
\centering
\includegraphics[width=0.7\textwidth]{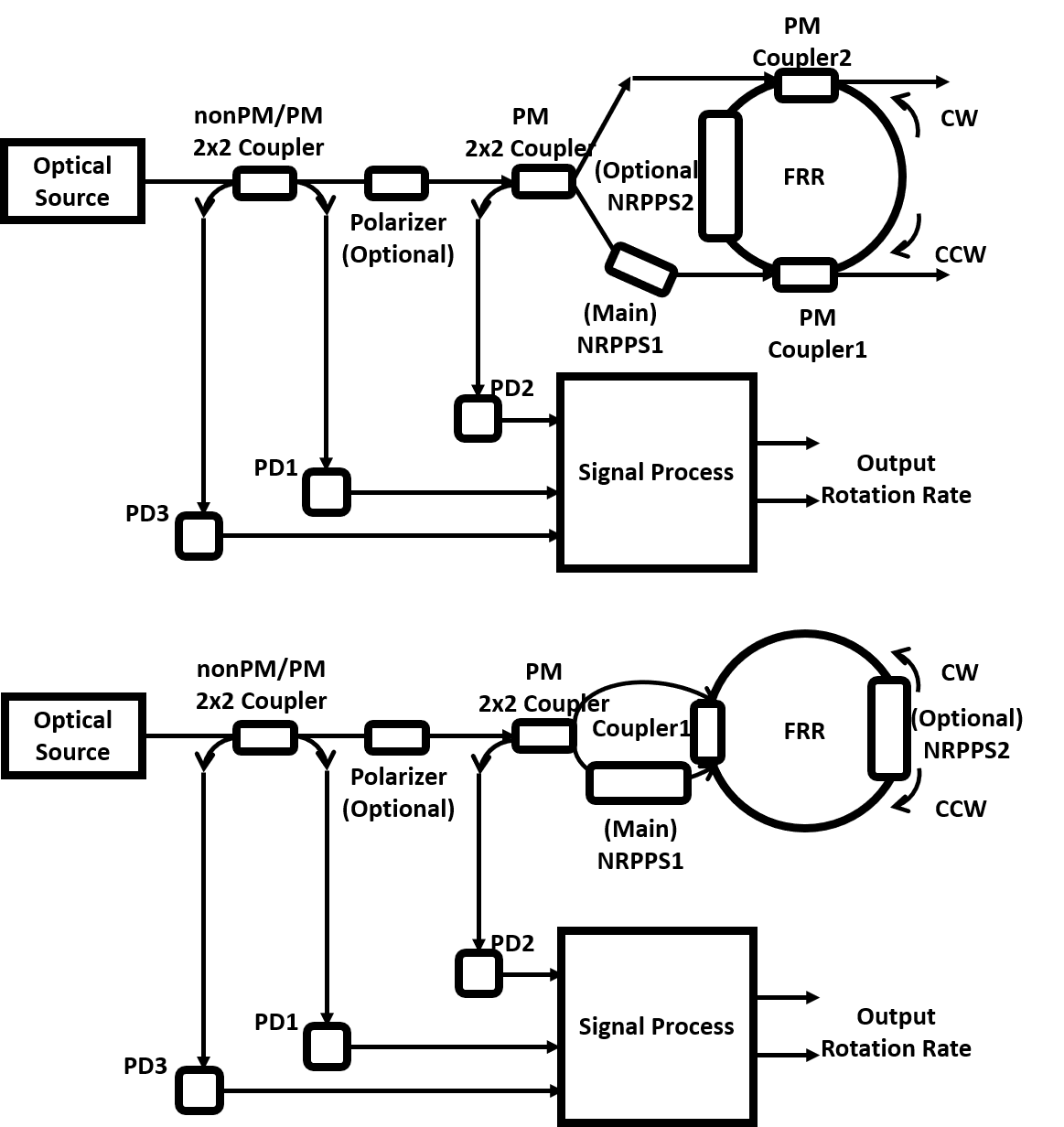}
\caption{\label{fig:NRPPS-RFOG_Setup} A Non-Reciprocal Polarization Phase Shifter based RFOG system (NRPPS-RFOG) with two-coupler FRR configuration (top), one-coupler FRR configuration (below).}
\end{figure}

The design of the RFOG employing ultra-simple white-light multibeam interferometry \cite{zhao2022navigation} is shown in Fig.~\ref{fig:RFOG_Setup}. A broadband light source, such as a superluminescent diode, is used in place of a narrow-linewidth laser. The light is divided into two paths and coupled into the fiber resonator via a multi-integrated optical chip (MIOC), which simultaneously functions as a 1×2 coupler, a polarizer, and a modulator. Within the resonator, formed by two fiber couplers (Coupler1 and Coupler2), multiple recirculations of the optical field generate sharp resonance modes that carry the Sagnac phase shift under rotation. The transmitted spectrum contains multiple cavity resonances excited by the broadband source, whose superposition yields a spectral interference pattern incorporating the resonance information of both clockwise (CW) and counter-clockwise (CCW) beams. By using white-light multibeam interferometry, the system extracts the relative frequency shift between the two propagation directions without the need for frequency locking, as the broadband excitation naturally spans the resonance spectrum. Nevertheless, to resolve the rotation direction, a modulation–demodulation scheme is still required through the MIOC, since passive interference alone cannot discriminate CW from CCW propagation. A simple fiber coupler and photodetector are sufficient to capture the interference signal, which is subsequently processed to recover both the magnitude and direction of rotation. As a conclusion, the setup converts the rotation-induced resonance splitting into a directly measurable signal while significantly reducing system complexity compared to traditional RFOG architectures.

\begin{figure}[!b]
\centering
\includegraphics[width=0.7\textwidth]{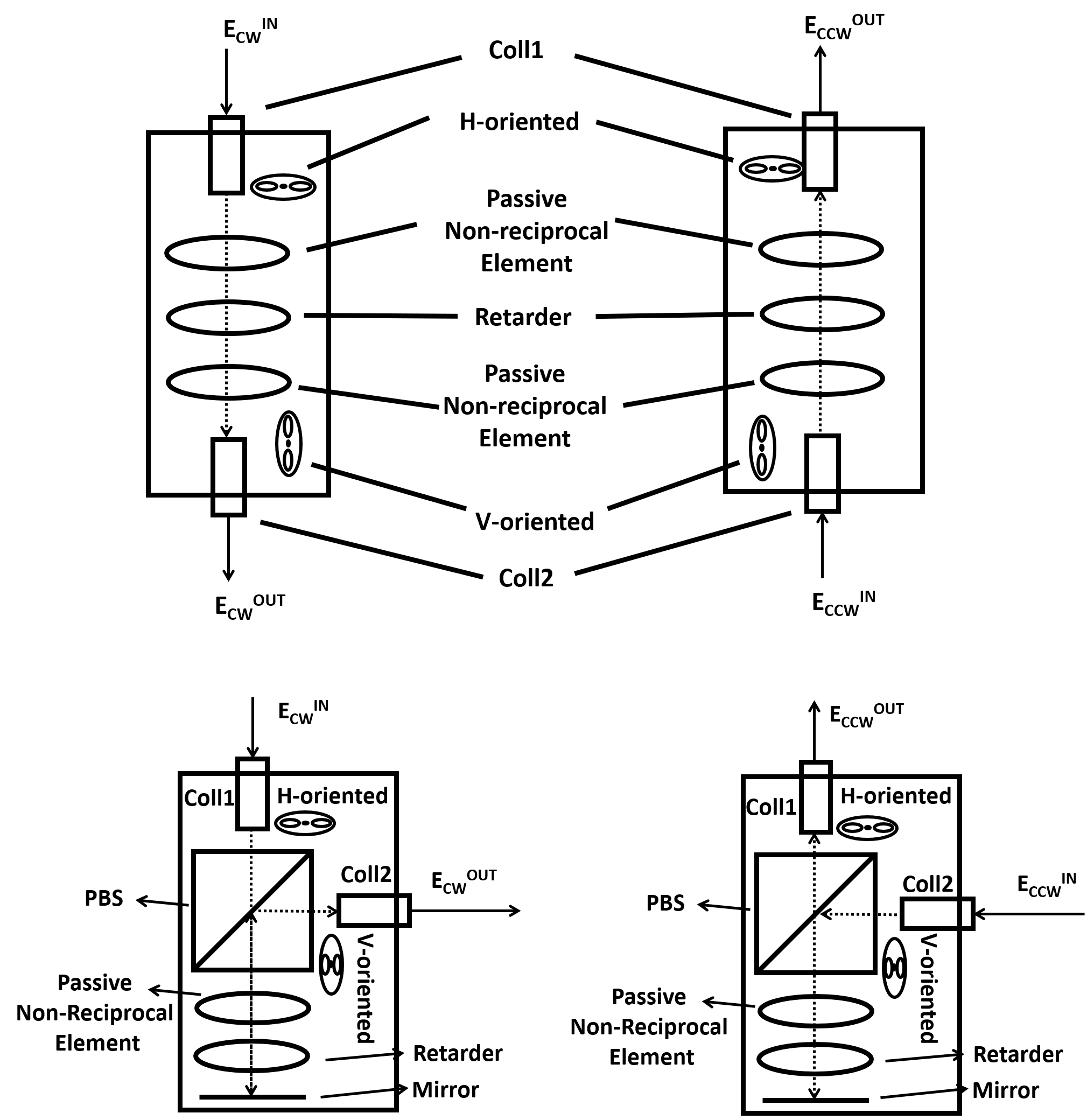}
\caption{\label{fig:NRPPSs_Design} Schematic of the Non-Reciprocal Polarization-Dependent Phase Shifter (NRPPS) showing the propagation of clockwise (CW) and counter-clockwise (CCW) beams: the transmission-type (top) and the reflection-type (bottom).}
\end{figure}

To further simplify the RFOG architecture, we propose the integration of a Non-Reciprocal Polarization-Dependent Phase Shifter (NRPPS) into the broadband resonant scheme, referred to here as the NRPPS-RFOG, as illustrated in Fig.~\ref{fig:NRPPS-RFOG_Setup}. In this work, we focus on the two-coupler FRR configuration of the NRPPS-RFOG, in order to exclude intensity contributions that do not originate from the resonant cavity itself. The NRPPS was originally introduced in the context of IFOGs as a passive technique to establish a stable quadrature bias without the need for active phase modulators such as the MIOC or piezoelectric devices \cite{akcaalan2025phase,akcaalan2025simple}. In this approach, non-reciprocal polarization rotation combined with birefringent elements generates a fixed phase bias between the counter-propagating beams. This mechanism enables the passive achievement of a stable $\pi/2$ quadrature point, thereby eliminating the necessity for modulation–demodulation loops and simultaneously reducing noise, complexity, and cost. When applied to the RFOG configuration, the NRPPS is placed within the resonant cavity to bias the CW and CCW resonance modes. This passive biasing not only provides a well-defined quadrature point but also maximizes sensitivity to the Sagnac phase shift. Furthermore, it overcomes a key limitation of earlier passive design based on 3×3 couplers \cite{ovchinnikov2023prototype}, in which the quadrature point deviates from $\pi/2$, leading to reduced sensitivity.

The NRPPS scheme is illustrated in Fig.~\ref{fig:NRPPSs_Design}. In the transmission-type configuration, the light passes through the retarder only once, requiring the element to be set as a quarter-wave plate. In the reflection-type configuration, the light passes through the same retarder twice, necessary of using an eight-wave plate. This design ensures that a stable $\pi/2$ quadrature point is consistently maintained between the CW and CCW beams. As the passive non-reciprocal element, a simple Faraday rotator can be used. A detailed Jones matrix analysis and the corresponding calculations of the NRPPS are provided in prior studies \cite{akcaalan2025phase,akcaalan2025simple}. To mitigate polarization-induced errors, a polarizer may be placed at each collimator in the transmission-type configuration; however, this method is not applicable to the reflection-type. For this reason, the transmission-type NRPPS is selected for integration into the RFOG design.

The operating principle of the NRPPS-RFOG is closely similar to that of the RFOG employing ultra-simple white-light multibeam interferometry \cite{zhao2022navigation,ovchinnikov2023prototype}. The electric fields of the CW and CCW beams, immediately following the second 2×2 PM coupler as seen in Fig.~\ref{fig:NRPPSs_Design}, can be expressed as:

\begin{eqnarray}
\centering
&&\label{eqnarray:1} E_{CW} = \frac{1}{\sqrt{2}}E_{0}e^{({i(\omega t+pi/2)})},\\
&&\label{eqnarray:2} E_{CCW} = \frac{1}{\sqrt{2}}E_{0}e^{({i(\omega t+0)})}
\end{eqnarray}

Here, $E_{0}$ represents the total electric field of the ASE at the input of the second 2×2 PM coupler, and the $\pi/2$ phase-shift difference arises from the intrinsic design of the coupler. This phase offset is canceled when the CW and CCW beams return to the same coupler from the fiber loop, owing to the symmetry of the system.

Lets first consider the CW direction. Before entering the first coupler named as coupler1 that forms the FRR section, the CW beam acquires a relative phase shift $\phi_{NRPSS}$ induced by the NRPPS. It then propagates to Coupler1, where an additional phase shift $\phi_{coupler1}$ arises due to the coupling ratio,

\begin{eqnarray}
\centering
&&\label{eqnarray:1} E_{CW} = \frac{1}{\sqrt{2}}E_{0}e^{({i(\omega t+pi/2+\phi_{NRPPS1})})}\sqrt{(1-R)}e^{({i(\omega t_1+\phi_{coupler1})})}
\end{eqnarray}

Here, $R$ imply the coupling ratio of Coupler1 in the FRR section, and $\phi_{coupler1}$ represents the additional phase shift introduced by Coupler1 due to its intrinsic properties. When an angular rotation rate $\Omega$ is present, a Sagnac phase difference arises between the CW and CCW beams, which can be expressed as:

\begin{eqnarray}
\centering
&&\label{eqnarray:1} \Delta \phi_{s} = \frac{8 \pi A N}{\lambda c }  \Omega 
\end{eqnarray}

Here, $A$ is the area of the FRR for one turn of the loop, $N$ is the number of turns of the coil, $\lambda$ is the central wavelength of the ASE source, and $c$ is the speed of light. The resulting electric field of the CW beam can then be expressed as:

\begin{eqnarray}
\centering
&&\label{eqnarray:1} E_{CW} = \frac{1}{\sqrt{2}}E_{0}e^{({i(\omega t+pi/2+\phi_{NRPPS1})})}\sqrt{(1-R)}e^{({i(\omega t_1+\phi_{coupler1}+\frac{\Delta\phi_s}{2}})}, \\
&&\label{eqnarray:2} \textrm{where}\ t_1 = \frac{Ln}{c}
\end{eqnarray}

Here, $L$ is the length of the FRR. The phase shift is defined as half of the total Sagnac phase shift, $+\frac{\Delta\phi_s}{2}$ for the CW beam and $-\frac{\Delta\phi_s}{2}$ for the CCW beam, so that the total phase difference remains unchanged. The CW beam then reaches the second coupler, indicating as Coupler2, which completes the FRR section and allows the beam to exit the resonator after a single circulation. For simplicity, the coupling ratio of Coupler2 is taken to be identical to that of Coupler1. Upon exiting Coupler2, the CW beam acquires an additional phase shift $\phi_{coupler2}$ due to the coupler’s intrinsic properties, similar to $\phi_{coupler1}$ from Coupler1. The resulting electric field of the CW beam can then be expressed as:

\begin{eqnarray}
\centering
&&\label{eqnarray:1} E_{CW} = \frac{1}{\sqrt{2}}E_{0}e^{({i(\omega t+pi/2+\phi_{NRPPS1})})}(1-R)e^{({i(\omega t_1+2\phi_{coupler1}+\frac{\Delta\phi_s}{2}}))}
\end{eqnarray}

The CW beam then returns to the 2x2 PM coupler back. The output ports of this coupler will be analyzed subsequently to demonstrate the operation of the design at the $\pi/2$ and $3\pi/2$ quadrature points.

We now consider the CCW beam. The CCW beam first reaches Coupler2, getting a phase shift $\phi_{coupler2} = \phi_{coupler1}$. It then enters the FRR, experiencing a Sagnac-induced phase shift of $-\frac{\Delta\phi_s}{2}$, and proceeds to Coupler1. Assuming a single circulation within the FRR, the CCW beam exits the resonator via Coupler1, getting an additional phase shift $\phi_{coupler1}$ with a coupling ratio of $1-R$. Subsequently, the beam passes through the NRPPS unit without any additional relative phase and finally reaches the 2x2 PM coupler.

\begin{eqnarray}
\centering
&&\label{eqnarray:1} E_{CCW} = \frac{1}{\sqrt{2}}E_{0}e^{({i(\omega t+0+0)})}(1-R)e^{({i(\omega t_1+2\phi_{coupler1}-\frac{\Delta\phi_s}{2}}))}
\end{eqnarray}

If the beams complete an additional circulation within the FRR loop, both the CW and CCW beams get an additional Sagnac phase shift with the coupling ratio $R$.

\begin{eqnarray}
\centering
&&\label{eqnarray:1} E_{CW} = \frac{1}{\sqrt{2}}E_{0}e^{({i(\omega t+pi/2+\phi_{NRPPS1})})}(1-R)e^{(i2\phi_{coupler1})}(Re^{({i(\omega t_1+\frac{\Delta\phi_s}{2}}))}+R^2e^{({2i(\omega t_1+\frac{\Delta\phi_s}{2}}))}), \\
&&\label{eqnarray:2} E_{CCW} = \frac{1}{\sqrt{2}}E_{0}e^{({i(\omega t+0+0)})}(1-R)e^{(i2\phi_{coupler1})}(Re^{({i(\omega t_1-\frac{\Delta\phi_s}{2}}))}+R^2e^{({2i(\omega t_1-\frac{\Delta\phi_s}{2}}))})
\end{eqnarray}

As shown in the following formula, each circulation introduces a Sagnac phase shift weighted by the coupling ratio $R$, assuming that the CW and CCW beams complete the same number of turns, such that:

\begin{eqnarray}
\centering
\textrm{for CW beam}\ 
&&\label{eqnarray:1} \sum_{p=0}^{\infty} R^{p}e^{({i(\omega t_1+\frac{\Delta\phi_s}{2}}))p},\\
\textrm{for CCW beam}\ 
&&\label{eqnarray:2} \sum_{p=0}^{\infty} R^{p}e^{({i(\omega t_1-\frac{\Delta\phi_s}{2}}))p}
\end{eqnarray}

Then, the general expression can be written up to the 2x2 PM coupler as:

\begin{eqnarray}
\centering
&&\label{eqnarray:1} E_{CW} = \frac{1}{\sqrt{2}}E_{0}e^{({i(\omega t+pi/2+\phi_{NRPPS1})})}(1-R)e^{(i2\phi_{coupler1})}\sum_{p=0}^{\infty} R^{p}e^{({i(\omega t_1+\frac{\Delta\phi_s}{2}}))p}, \\
&&\label{eqnarray:2} E_{CCW} = \frac{1}{\sqrt{2}}E_{0}e^{({i(\omega t+0+0)})}(1-R)e^{(i2\phi_{coupler1})}\sum_{p=0}^{\infty} R^{p}e^{({i(\omega t_1-\frac{\Delta\phi_s}{2}}))p}
\end{eqnarray}

When the CW and CCW beams pass through the 2x2 PM coupler, their e-fields are superposed. At PD2, due to the symmetry of the coupler, the CW beam acquires an additional $\pi/2$ phase shift, whereas the CCW beam reaches PD2 without any extra phase contribution. The resulting CW and CCW e-fields at PD2 can thus be expressed as:

\begin{eqnarray}
\centering
&&\label{eqnarray:1} E_{CW-PD2} = \frac{1}{\sqrt{2}}E_{0}e^{({i(\omega t+pi/2+\phi_{NRPPS1}+pi/2)})}(1-R)e^{(i2\phi_{coupler1})}\sum_{p=0}^{\infty} R^{p}e^{({i(\omega t_1+\frac{\Delta\phi_s}{2}}))p}, \\
&&\label{eqnarray:2} E_{CCW-PD2} = \frac{1}{\sqrt{2}}E_{0}e^{({i(\omega t+0+0+0)})}(1-R)e^{(i2\phi_{coupler1})}\sum_{p=0}^{\infty} R^{p}e^{({i(\omega t_1-\frac{\Delta\phi_s}{2}}))p}
\end{eqnarray}

For PD3, the phase shift introduced by the 2x2 PM coupler is the reverse of that at PD2. In addition, due to the half transmission from the second arm, the collected CW and CCW fields at PD3 can be expressed as:

\begin{eqnarray}
\centering
&&\label{eqnarray:1} E_{CW-PD3} = \frac{1}{\sqrt{2}}\frac{1}{\sqrt{2}}E_{0}e^{({i(\omega t+pi/2+\phi_{NRPPS1}+0)})}(1-R)e^{(i2\phi_{coupler1})}\sum_{p=0}^{\infty} R^{p}e^{({i(\omega t_1+\frac{\Delta\phi_s}{2}}))p}, \\
&&\label{eqnarray:2} E_{CCW-PD3} = \frac{1}{\sqrt{2}}\frac{1}{\sqrt{2}}E_{0}e^{({i(\omega t+0+0+pi/2)})}(1-R)e^{(i2\phi_{coupler1})}\sum_{p=0}^{\infty} R^{p}e^{({i(\omega t_1-\frac{\Delta\phi_s}{2}}))p}
\end{eqnarray}

Next, the interference signals corresponding to PD2 and PD3 are derived by evaluating the optical intensities obtained from the superposed CW and CCW e-fields.

\begin{eqnarray}
\centering
&&\label{eqnarray:1} E_{PD2} = \frac{1}{\sqrt{2}}E_{0}e^{{i(\omega t)}}(1-R)e^{(i2\phi_{coupler1})}\sum_{p=0}^{\infty} R^{p}e^{i(\omega t_1)p}[e^{i(\pi+\phi_{NRPPS1}+\frac{\Delta\phi_s}{2}p)}+e^{i(0-\frac{\Delta\phi_s}{2}p)} \\
&&\label{eqnarray:2} E_{PD3} = \frac{1}{\sqrt{2}}\frac{1}{\sqrt{2}}E_{0}e^{{i(\omega t)}}(1-R)e^{(i2\phi_{coupler1})}\sum_{p=0}^{\infty} R^{p}e^{i(\omega t_1)p}[e^{i(\frac{\pi}{2}+\phi_{NRPPS1}+\frac{\Delta\phi_s}{2}p)}+e^{i(\frac{\pi}{2}-\frac{\Delta\phi_s}{2}p)}
\end{eqnarray}

By applying the standard exponential-sum representation, 

\begin{eqnarray}
\centering
&&\label{eqnarray:1} e^{i\phi}+e^{i\beta} = 2\cos{((\phi-\beta)/2)}e^{i(\phi+\beta)/2}
\end{eqnarray}

the e-fields on PD2 and PD3 can be reformulated as follows:

\begin{eqnarray}
\centering
&&\label{eqnarray:1} E_{PD2} = C\sum_{p=0}^{\infty} R^{p}e^{i(\omega t_1)p}cos(\frac{\Delta\phi_s}{2}p+\frac{\pi+\phi_{NRPPS1}}{2}), \\
&&\label{eqnarray:2} E_{PD3} = \frac{C}{\sqrt2}\sum_{p=0}^{\infty} R^{p}e^{i(\omega t_1)p}cos(\frac{\Delta\phi_s}{2}p+\frac{\phi_{NRPPS1}}{2}),\\
\textrm{where}\ 
&&\label{eqnarray:3}  C = \frac{2}{\sqrt{2}}E_{0}e^{{i(\omega t)}}(1-R)e^{(i2\phi_{coupler1})}e^{i(\frac{\pi+\phi_{NRPPS1}}{2})}
\end{eqnarray}

The optical power detected at PD2 and PD3 is obtained as the conjugate product of the corresponding e-fields and integrated over the entire frequency spectrum of the fields, similar as \cite{ovchinnikov2023prototype}, such that:

\begin{eqnarray}
\centering
&&\label{eqnarray:1} P(f)_{PD2} = \int^{f2}_{f1}E(f)_{PD2}E(f)^*_{PD2}df, \\
&&\label{eqnarray:2} P(f)_{PD3} = \int^{f2}_{f1}E(f)_{PD3}E(f)^*_{PD3}df
\end{eqnarray}

Here, the angular frequency is defined as $\omega = 2\pi f$. The resulting signals at PD2 and PD3 can then be expressed as:

\begin{figure}[!t]
\centering
\includegraphics[width=0.45\textwidth]{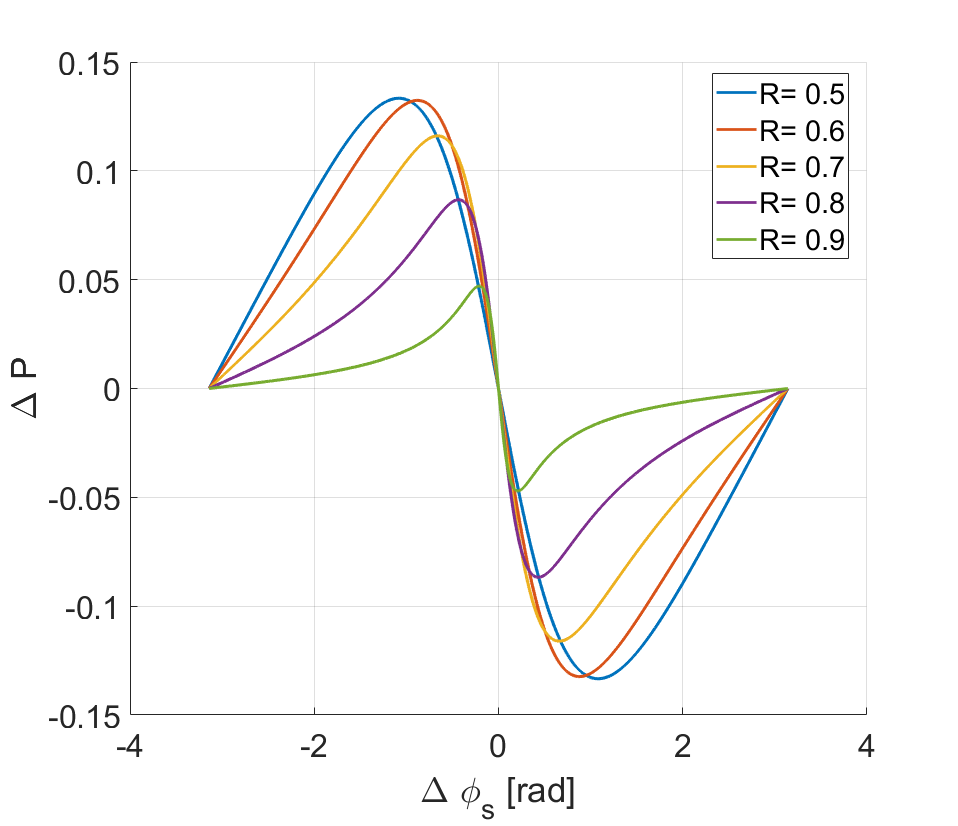}
\includegraphics[width=0.45\textwidth]{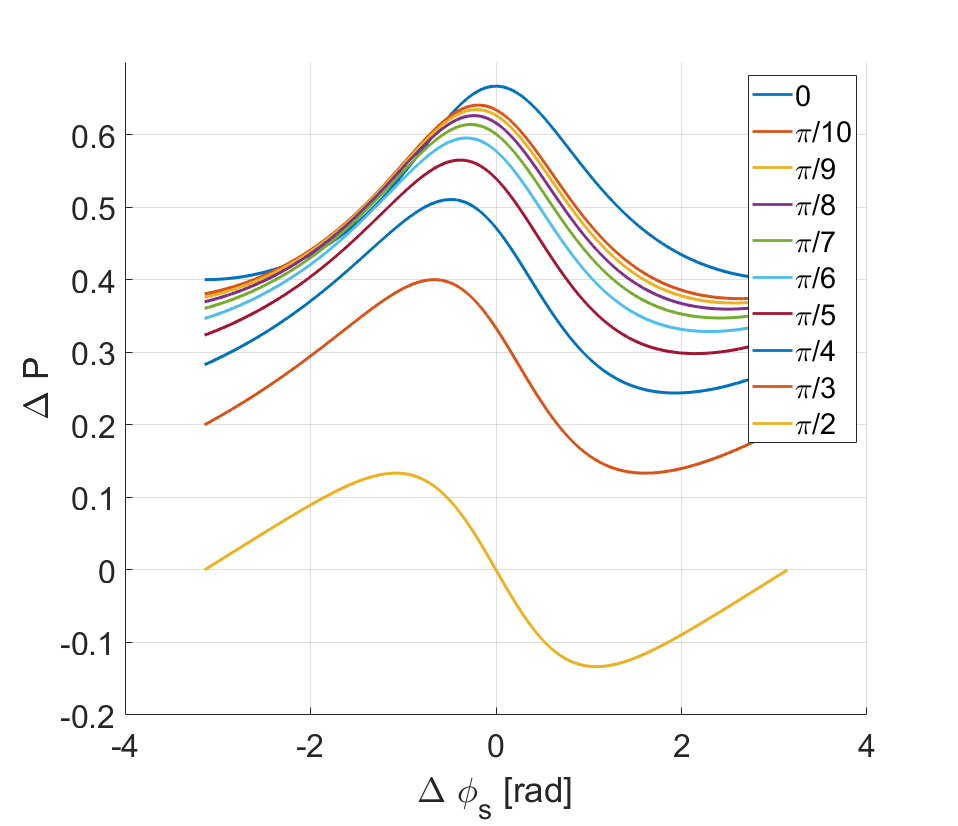}
\caption{\label{fig:DelPOptimum} $\Delta P$ variation over different $R$ values for $\phi_{NRPPS1} = \pi/2$ (left), $\Delta P$ variation over different $\phi_{NRPPS1}$ values for $ R = 0.5$ (right)}
\end{figure}

\begin{figure}[!b]
\centering
\includegraphics[width=0.45\textwidth]{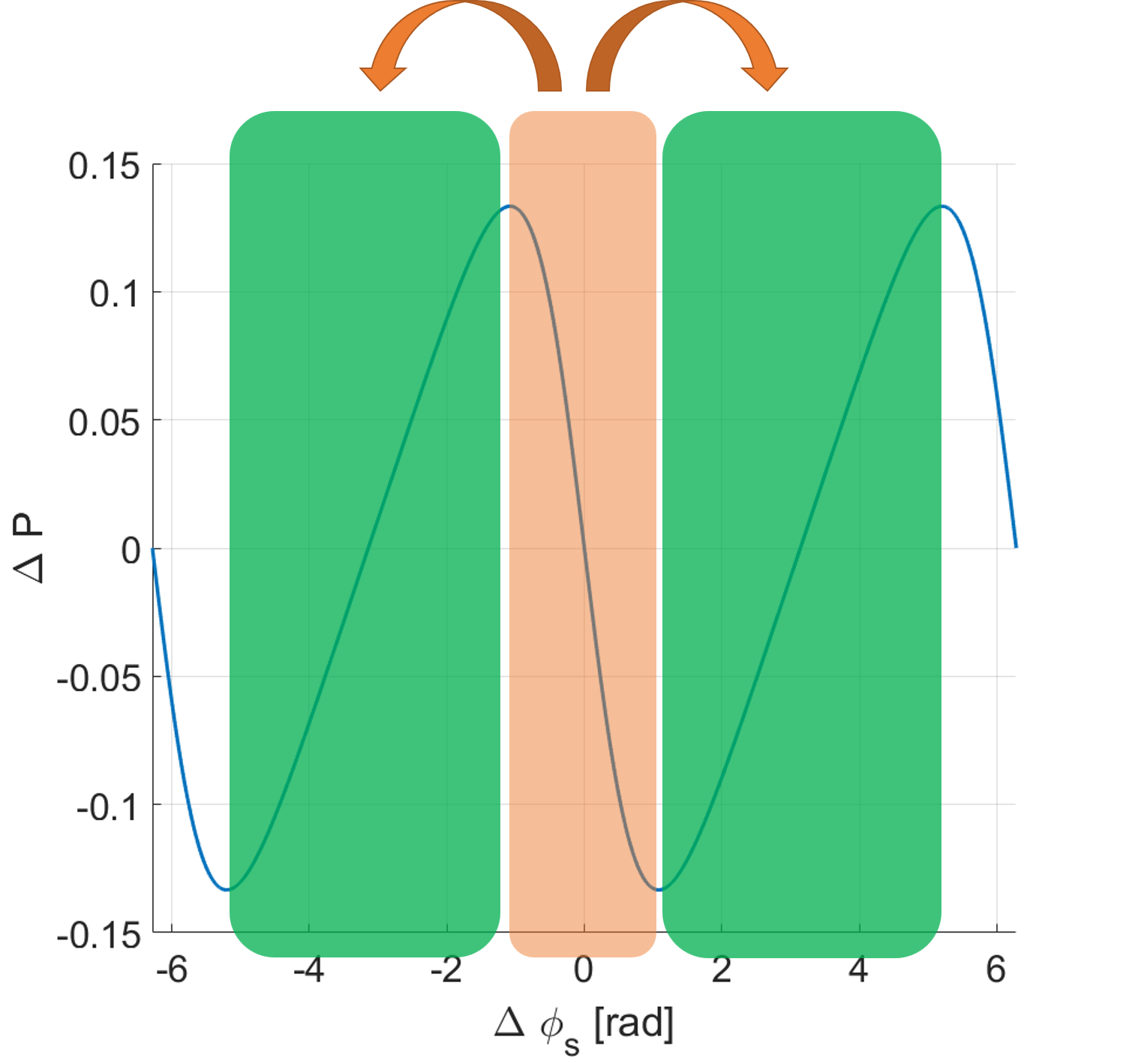}
\includegraphics[width=0.45\textwidth]{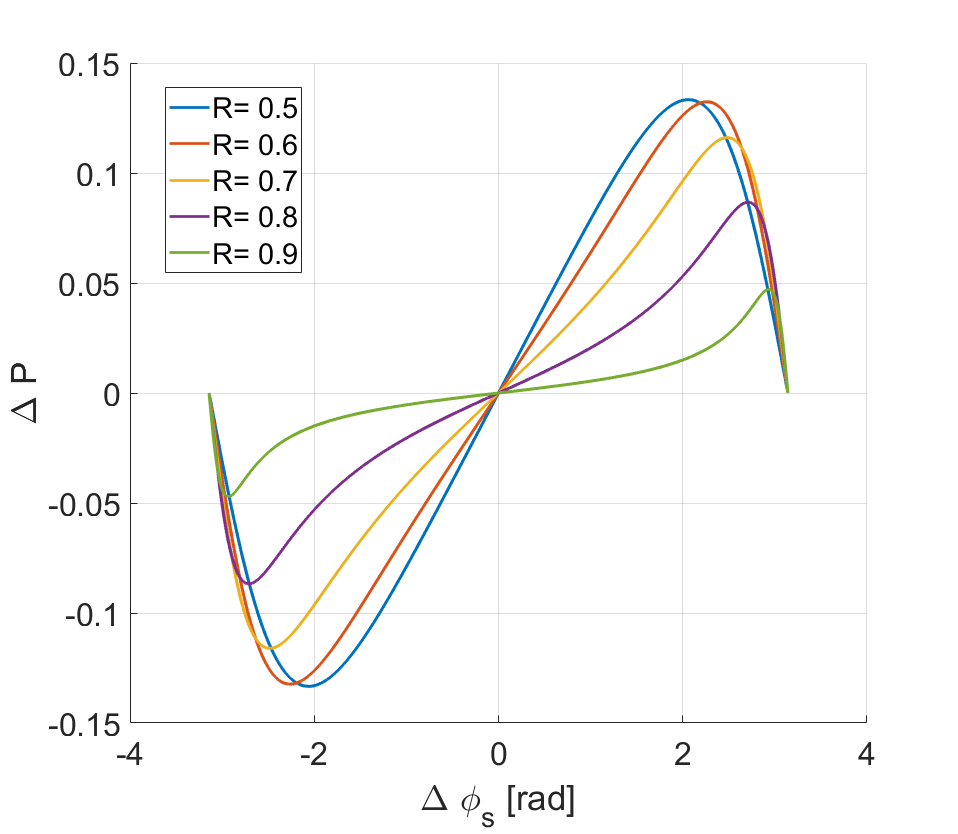}
\caption{\label{fig:DelPOver2PI} $\Delta P$ variation over $-2\pi$ to $2\pi$ for $\phi_{NRPPS1} = \pi/2$ and $ R = 0.5$ values (left), $\Delta P$ variation over different $R$ values for $\phi_{NRPPS1} = \pi/2$ and $\phi_{opt.NRPPS} = \pi$}
\end{figure}

\begin{eqnarray}
\centering
&&\label{eqnarray:1} P(f)_{PD2} =  C^2\sum_{m=0}^{\infty} R^{2m}cos^2(\frac{\Delta\phi_s}{2}(m)+\frac{\pi+\phi_{NRPPS1}}{2}), \\
&&\label{eqnarray:2} P(f)_{PD3} = \frac{C^2}{2}\sum_{m=0}^{\infty} R^{2m}cos^2(\frac{\Delta\phi_s}{2}(m)+\frac{\phi_{NRPPS1}}{2}),\\
\textrm{where}\ 
&&\label{eqnarray:3}  C^2 = 2P_0(1-R)^2
\end{eqnarray}

By employing the identities $\cos^2{x} = \frac{1+\cos{2x}}{2}$ and $\cos{A}-\cos{B} = -2\sin{\frac{A+B}{2}}\sin{\frac{A-B}{2}}$, the power difference at the detector can be expressed as:

\begin{eqnarray}
\centering
&&\label{eqnarray:1} \Delta P = P_{PD2} - 2P_{PD3} =  G\sum_{m=0}^{\infty} R^{2m}\sin{(\Delta\phi_s m+\frac{\pi+2\phi_{NRPPS1}}{2}}),\\
\textrm{where}\ 
&&\label{eqnarray:2}  G = C^2 = 2P_0(1-R)^2
\end{eqnarray}

By applying the standard closed-form expression for the geometric–sine series for $|R^2|<1$,

\begin{eqnarray}
\centering
&&\label{eqnarray:1} G\sum_{m=0}^{\infty} R^{2m}\sin{(Am+B)} = \frac{\sin{B}-R^2\sin{(B-A)}}{1-2R^2\cos{A}+R^4},
\end{eqnarray}

\begin{figure}[!b]
\centering
\includegraphics[width=0.9\textwidth]{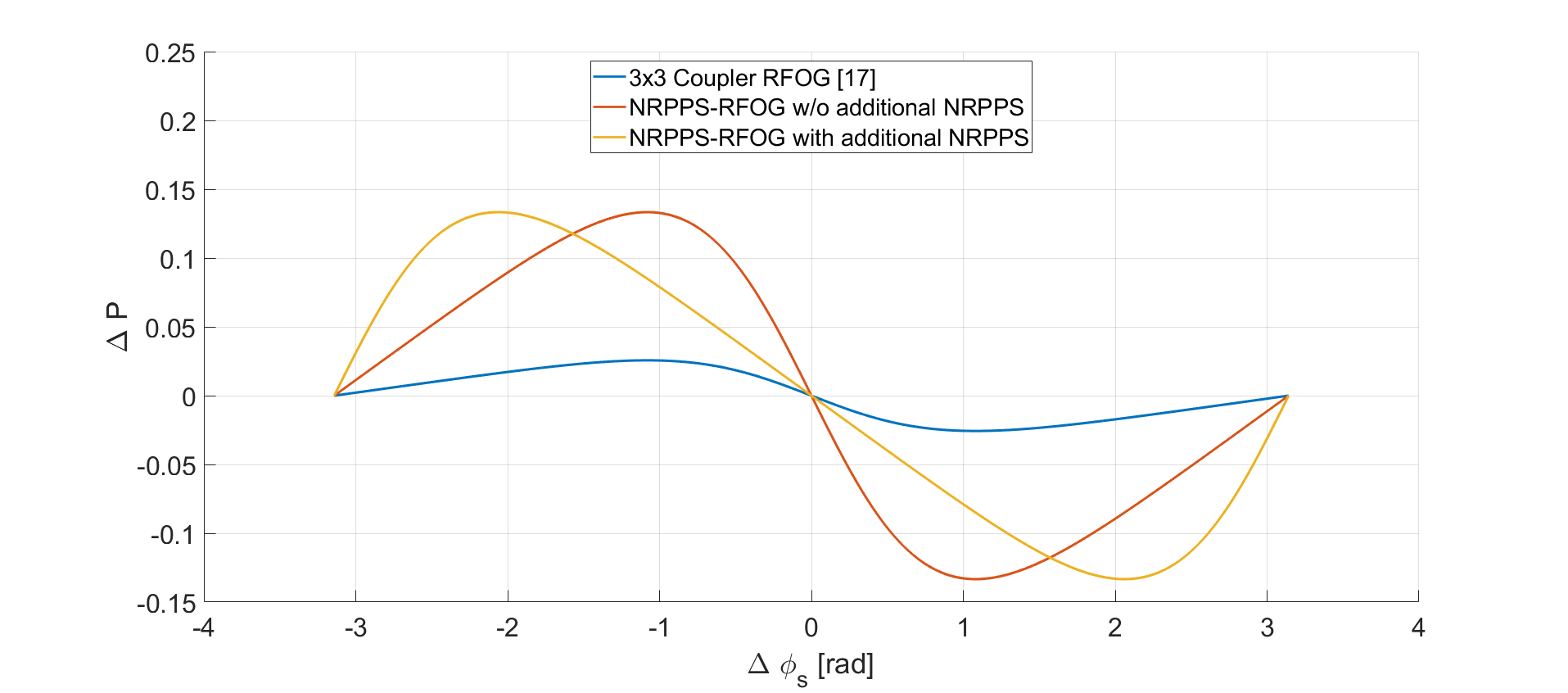}
\caption{\label{fig:DelPComparison} $\Delta P$ comparison over $-\pi$ to $\pi$ angular rotation shift for the passive 3x3 coupler RFOG design \cite{ovchinnikov2023prototype} and NRPPS-RFOG design at $\phi_{NRPPS1} = \pi/2$ and $\phi_{opt.NRPPS} = \pi$ for $ R = 0.5$ }
\end{figure}

the $\Delta P$ which represents the power difference of PD2 and PD3 can be written as:

\begin{eqnarray}
\centering
&&\label{eqnarray:1} \Delta P = \sin{(\frac{\pi+2\phi_{NRPPS1}}{2})}\frac{2P_0(1-R)^2}{1-2R^2\cos{{(\Delta\phi_s )}}+R^4}-\frac{2P_0(1-R)^2R^2\sin{( \frac{\pi+2\phi_{NRPPS1}}{2} - \Delta\phi_s ) }}{1-2R^2\cos{{(\Delta\phi_s )}}+R^4},
\end{eqnarray}

As shown in Eq.33, when $\phi_{NRPPS1} = (2n+1)\pi/2$ with $n \geq 1$ being an integer, the power difference $\Delta P$ simplifies to:

\begin{eqnarray}
\centering
&&\label{eqnarray:1} \Delta P = \frac{2P_0(1-R)^2R^2\sin{(\Delta\phi_s ) }}{1-2R^2\cos{{(\Delta\phi_s )}}+R^4}
\end{eqnarray}

Depending on the coupling ratios of Coupler1 and Coupler2 (assumed identical), $R$, the detector exhibits different responses, as illustrated in Fig.~\ref{fig:DelPOptimum} (left). The selection of $\phi_{NRPPS1} = (2n+1)\pi/2$ serves not only to eliminate the first term in Eq.33 and achieve a more linear response, but also to maximize the sensitivity, as illustrated in Fig.~\ref{fig:DelPOptimum} (right). The highest sensitivity is observed for a coupling ratio of approximately $R = 0.5$.

While analyzing $\Delta P$ as a function of angular rotation, an asymmetry was observed as the angular rotation range increases. As shown in Fig.~\ref{fig:DelPOver2PI}, by shifting the readout region toward the left (orange) or right (green) section from the central point, the measurable angular rotation range can be extended. Although this adjustment reduces the sensitivity for whole range, the effective measurement range nearly doubles. To implement this shift, an additional NRPPS section, denoted as NRPPS2, is incorporated into the setup as seen in Fig.~\ref{fig:NRPPS-RFOG_Setup}. When NRPPS2 is included in the formulation, the resulting $\Delta P$ becomes:

\begin{eqnarray}
\centering
&&\label{eqnarray:1} \Delta P = \sin{(\frac{\pi+2\phi_{NRPPS1}}{2})}\frac{2P_0(1-R)^2}{1-2R^2\cos{{(\Delta\phi_s +\phi_{NRPPS2})}}+R^4} 
\nonumber\\
&& -\frac{2P_0(1-R)^2R^2\sin{( \frac{\pi+2\phi_{NRPPS1}}{2} - \Delta\phi_s -\phi_{NRPPS2} ) }}{1-2R^2\cos{{(\Delta\phi_s +\phi_{NRPPS2} )}}+R^4},
\end{eqnarray}

For longer FRR sections, while the sensitivity increases, the measurable range decreases; the inclusion of this additional phase shift helps to mitigate this limitation.

When compared with the previously reported passive RFOG design based on a 3×3 coupler \cite{ovchinnikov2023prototype}, the proposed approach demonstrates an improvement of approximately fivefold in sensitivity, as shown in Fig.~\ref{fig:DelPComparison}. Furthermore, with the incorporation of an additional NRPPS section, the measurement range is extended by a factor of two, while still maintaining a sensitivity enhancement of about 2.5 times, as illustrated in Fig.~\ref{fig:DelPComparison}.

\section{Conclusion}
In conclusion, we have presented a novel passive RFOG design that achieves two true quadrature points at $\pi/2$ and $3\pi/2$, enabling the determination of angular rotation with maximum sensitivity. By integrating a Non-Reciprocal Polarization-Dependent Phase Shifter (NRPPS) into the broadband RFOG configuration, the system operates passively, eliminating the need for active modulation–demodulation while maintaining high sensitivity. Theoretical analysis demonstrates that the proposed approach preserves the advantages of resonant enhancement, allowing shorter fiber lengths without sacrificing performance. Additionally, the use of a broadband light source prevent the requirement for precise frequency locking, simplifying system architecture and reducing complexity. Parametric studies indicate that the coupling ratio and NRPPS phase shift critically influence the sensitivity and linearity of the detector output. Furthermore, the introduction of a secondary NRPPS section provides a mechanism to extend the measurable angular rotation range, mitigating limitations associated with longer fiber lengths. Overall, the proposed passive NRPPS-RFOG represents a promising approach toward compact, high-sensitivity, and cost-effective resonant fiber-optic gyroscopes suitable for navigation-grade applications.

\section*{Acknowledgments}
A provisional patent application has been filed for the technology described in this paper. O.A. conducted the simulations, analyzed the results and supervised the manuscript. M.G.A. reviewed the manuscript.

\bibliographystyle{unsrt}
\bibliography{NRPPS-RFOG}

\end{document}